\documentclass[11pt, a4paper]{article} 
\usepackage{graphicx}
\usepackage[left=3cm, right=3cm, top=2cm]{geometry}
\begin{document}
\title{A New Physical Model of Pairing Mechanism in Superconductors:
Could the Electron itself be treated as a Composite Particle to Achieve Room Temperature Superconductor?}
\author{Zuhair. M. Hejazi, Iskra B. Hejazi\thanks{Z. M Hejazi is currently with the Department of Communication Engineering, Yarmouk University, Jordan;
I.B Hejazi is my late wife}}
\date{}
\maketitle

\begin{abstract}
The physical pictures of the electron pairing structure and pairing mechanisms in superconductors are reviewed.  An initial idea for a new physical picture of the origin and nature of the pairing is proposed. The idea is based on the assumption that the electron is no longer a single fundamental but a composite particle. This property is hidden in the normal state. How a natural pairing could occur in the superconducting state and the processes closely related to this change inside the atom are developed in a new physical picture with new insight(although it needs verification and real evidence for now). An attempt, to show that a zero resistance to a direct current and Josephson effects could be used as example evidences for this assumption, is presented by means of this new insight in general schematical analogy. A possible new research direction, hopefully to achieve room temperature superconductors, is suggested as a consequence.
Nowadays, with the great advancements of facilities used for particle physics, quantum mechanics and applied superconductivity, there is a great opportunity to observe, fabricate and test and possibly prove/disprove this new insight
\end{abstract}


\section{Introduction}

Regardless whether the superconductors are metallic, ceramic or even organic, low, moderate or high  critical temperature, the absolute consensus in the theory of superconductivity is that there is a kind of pairing mechanism between electrons. This pairing is responsible for the superconducting state and is the heart of all proposed theories to date.  The picture of the internal nature of this pairing is thus the most controversial problem that still a mystery.
It has direct relevance to the RF properties of high $T_c$ superconductors (HTS) and to the behaviour of high $T_c$ Josephson junctions and others of great practical importance.  Understanding the origin of the pairing properly will, with no doubt, help us to find the way for producing more stable, simpler, reliable higher $T_c$ materials and even, hopefully, room temperature superconductors.
It looked as the lattice played no role in the superconducting mechanism as X-ray studies showed that there was no change in either the symmetry or the spacing of the lattice  when superconductivity
occurred \cite{Smart}.  Nearly 40 years after the discovery of superconductivity in 1911 \cite{Onnes} and 20 years after the discovery of Meissner effect in 1933 \cite{Meissn}, however an isotope effect was observed \cite{Maxw}: $T_c$ was found to depend on the isotopic mass M such that:
 $T_c \propto \frac{1}{\sqrt{M}}$. The frequency, $\nu$, of vibration of a diatomic molecule is known to be given as $\nu = \frac{1}{2 \pi} \sqrt{\frac{k}{m}}\;$ where $m$ is the reduced mass of the molecule, and $k$ is the force constant of the bond.  It is shown that a vibration changes frequency on isotopic substitution such that the frequency, $\nu$, is proportional to $\frac{1}{\sqrt{mass}}$.   This  is  what suggested to physicists that superconductivity was in some way related to the vibrational modes of an isolated molecule, the  quanta of lattice vibrations are  called  phonons.

It  was suggested  that  there  could  be  a  strong  phonon-electron interaction in a superconductor that leads to an attractive force between two electrons, strong enough, to overcome the coulomb repulsion
between them.  Based on this picture, Bardeen, Cooper and Schrieffer published their theory of superconductivity, well known as the BCS \cite{BCS} theory.  It is commonly accepted for conventional low temperature superconductors (LTS) where the electron-phonon mechanism is dominant. The discovery of (HTS) \cite{Bedmu} makes it uncertain to explain the phenomenon in the new materials.  Firstly, the net attraction resulting from the  lattice does not appear to be strong enough to account for the high $T_c$ values. Secondly, although there is evidence that pairs of electrons are still responsible for HTS, it is not  yet  clear  whether the interaction with each other is via a lattice vibration, because  the isotopic  substitution data is contradictory. These problems are discussed in detail in the literature \cite{Frolich}\cite{Philips89}\cite{BatlagFaltens87} \cite{Batlagetal87}\cite{LoyeGarcia88}\cite{AndersAbrah87}.

The common search is concentrated to find a different mechanism for HTS. Enormous number of theories and models have been proposed for the pairing mediated interactions and their mechanisms.

The phonon mechanism has been the best candidate for many years before the discovery of HTS. Much about this mechanism and the constraints facing it before and after the BCS theory is discussed in \cite{Born}\cite{Cooper}\cite{Grimv} \cite{Allen}\cite{Eliash}\cite{Kresin2} \cite{Millan}.
Intensive search for other pairing mechanisms or other excitations rather than phonons has been motivated to explain the phenomenon.

The idea of electronic mechanism is enhanced by the fact that the electronic energy scale is much higher than the ionic. This factor may be responsible for the hope that high $T_c$ may be achieved in this way
\cite{Little1}\cite{Little2}\cite{Areview} \cite{YangPRL89}\cite{JandeB}\cite{PWASci87}\cite{Ruvalds1} \cite{Ruvalds2}  \cite{BarYam}\cite{HolcPR94}.

Essentially, theorists are trying to find an additional or other types of mediators between the electrons rather than the lattice (envisioned by the BCS theory for LTS). These mediators should transform a repulsive force into an attractive one to meet the new and unexpected challenge emerged with the discovery of the new HTS.  A magnetic and mixed mechanisms were also proposed in several models. It is beyond the scope of an article to capture all the pertinent proposals. Excellent reviews about the most recent theoretical trends and progress in HTS are found in \cite{Philips89}\cite{PWAPT91}\cite{BetlagPT91}\cite{BeasleyApp95}\cite{JRuvalds96}

 The issue of the symmetry of the pairing state (s-wave, d-wave or something else) is still controversial.  Basically this is a question of the internal structure of the electron pairs that could hide the puzzle of superconductivity mechanism.  Is this structure the same or different from that in conventional superconductors?

In conventional s-wave superconductors the pairs are believed to be formed by electrons with opposite spin and momentum.  In the d- wave picture, the members of a pair can be thought of as in counter circulating  orbits.  The wave function of the later changes sign which is equivalent to a phase change of $\pi$ \cite{BeasleyApp95}.  The phase shift predicted for the $d_{x^2-y^2}$ was really observed \cite{WollPRL93}.  If the pair potential is positive in one direction and negative in another, a phase shift of $\pi$ appears that is reflected in the flux dependence of the supercurrent \cite{TanakaPRL94}. Recent details reflecting the issue of the symmetry of the pairs of HTS can be found in \cite{ChaudYuPRL94}\cite{XuPRL94}\cite{YuPRL76}\cite{XuPRL95}\cite{DahmPRL95} \cite{WollPRL95}\cite{KellPRL94}.

Other Novel models for the origin of the pairing interaction between the carriers based on magnetic effects is the spin bag proposed by R.  Schrieffer \cite{PWAPT91} \cite{SchrPRB39}. It is a version of the magnetic polaron, in which one carrier locally perturbs the antiferromagnetic order by its own spin, forming a spin bag.  A second carrier within the coherence length of this local distortion experiences an attractive interaction by sharing the bag.
Schrieffer has emphasized \cite{SchrLTP95} that vertex corrections should be included in the study of spin-fluctuation pairing mechanisms, and the spin bag conclusions also require such further work.

 In  another novel model, P.  Anderson \cite{PWASci87} \cite{AndPRB39} proposed that magnetic fluctuations are responsible for the pairing and suggested that a novel type of ground state  consisting of singlet pairs may form a quantum spin liquid rather than the Fermi liquid picture.  The Fermi surface, however,  is not the surface of a Fermi sea of quasielectrons; rather \cite{PWAPT91} it is the surface of neutral, spin-carrying fermions called ``spinons''.  The ``spinons'' do not carry charge and so a second branch of the excitation spectrum is invented, the ``holons'', which are charged spinless objects.  The only objects that can move coherently from one plane to another ($CuO_2\;$ planes) are real electrons, but these break up incoherently into holons and spinons when they arrive \cite{PWAPT91} \cite{AndPRB42}. He saw, however, that the real difficulty is that one must use as carriers the same electrons that are participating in making an antiferromagnetically correlated spin structure.

Although the mechanism could be understood for LTS, it is still a mystery for the new HTS where the cause of the pairing is of great interest. Researchers from Daresbury and universities of Bristol in the UK and Stuttgart in Germany are using a method of calculating the quantum mechanical properties of solids, to identify the ``location'' of the Cooper pairs within the cristal lattice of the superconductor. In parallel, researchers at Stanford, Argonne and Wisconsin in the USA have been investigating the Fermi surface of superconductors. Angle-resolved photoemission experiments are hoped to reveal energy information that could hold the key to the pairing nature. It is believed that these groups will be able to provide information on the location of the electron pairing in a superconductor, giving insight into the origin of the pairing mechanism in HTS.

In the next sections, we will attempt to present an idea (assumption not yet taken into account) which would sound radical, but we will apply a well known and established experimental effects supporting this assumption as an example evidences.

\section{What if the Electron Itself is a Composite Particle?}

When we try to describe something complicated in quantum mechanical aspect, we first may use simple general analogies in classical mechanics.

\begin{figure}[ht]
\centering
\includegraphics[width=0.75\textwidth]{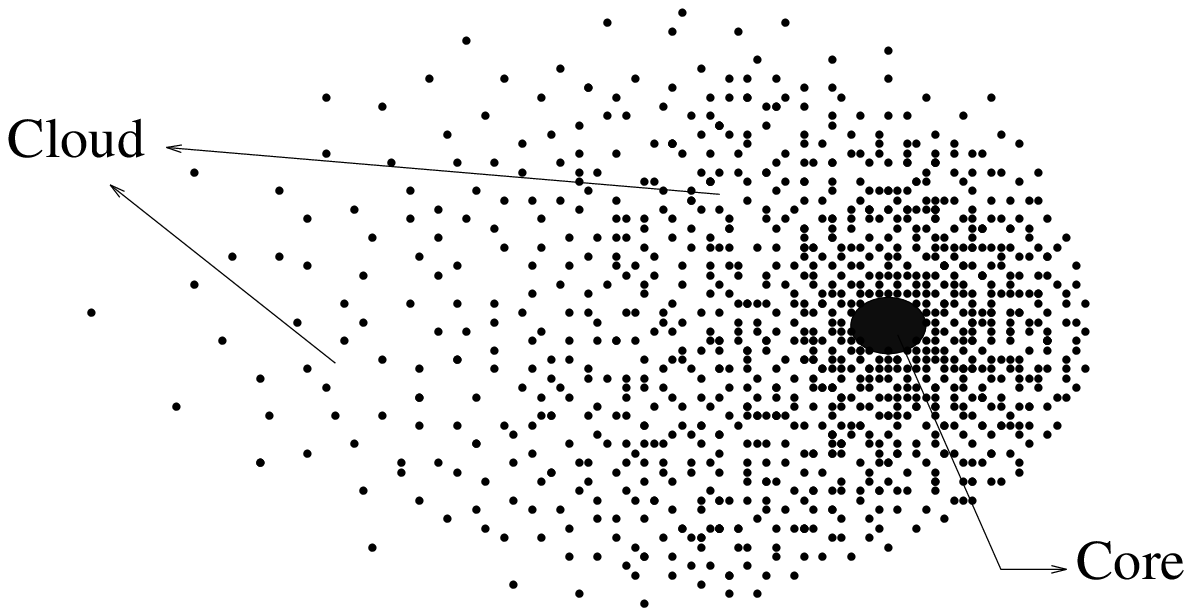}
\caption{The electron core where the cloud is condensed and bent around it.}
\label{cornorm}
\end{figure}

\begin{figure}[ht]
\centering
\includegraphics[width=0.75\textwidth]{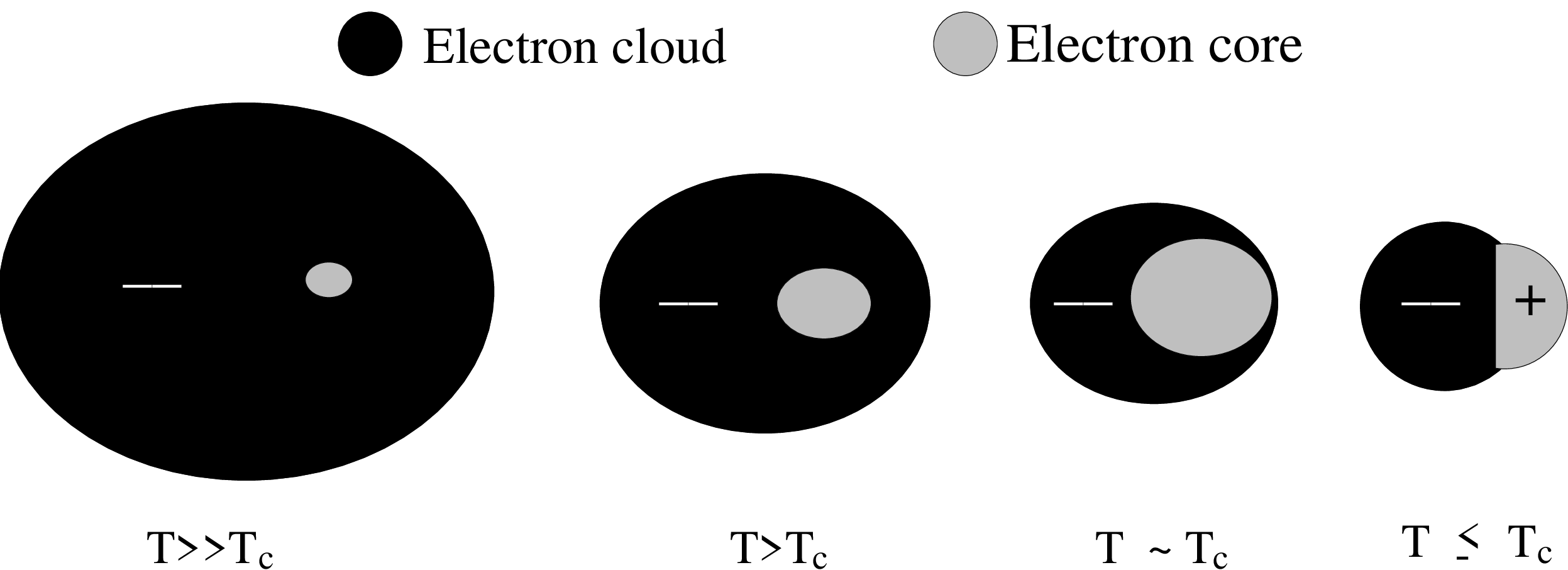}
\caption{Schematical presentation of forming the superpair from within the normal electron only.}
\label{minipair}
\end{figure}

The conventional picture of the electron structure itself, as a fundamental particle, may miss something important (maybe unobserveable in the normal state) which exhibits itself to the real world only in the superconducting state.
When a given material is driven to a superconducting state, the conduction electrons ought to have experienced a specific change.  The unusual observations concerning this specific change, would suggest that the electron might usefully no longer  be considered as the usual  simple single fundamental particle.  It is proposed that it be considered as a carrier with a kind of core inside it as shown in Fig. \ref{cornorm}.  In the normal state this core does not  exhibit any indications of its existence.  Transiting into the superconducting state, it is possible that the outer electron envelope (cloud) experiences a collapse (shrinkage) as a result of radiating (caused by the cooling process) its thermal energy. This leads the core under pressure  to ``swell up'' or ``break out''  as shown schematically in Fig. \ref{minipair}. At the same time this core begins to exhibit apparent properties of a ``positive charged particle'' because of its spin, which is opposite to the cloud.

\begin{figure}[ht]
\centering
\includegraphics[width=0.6\textwidth]{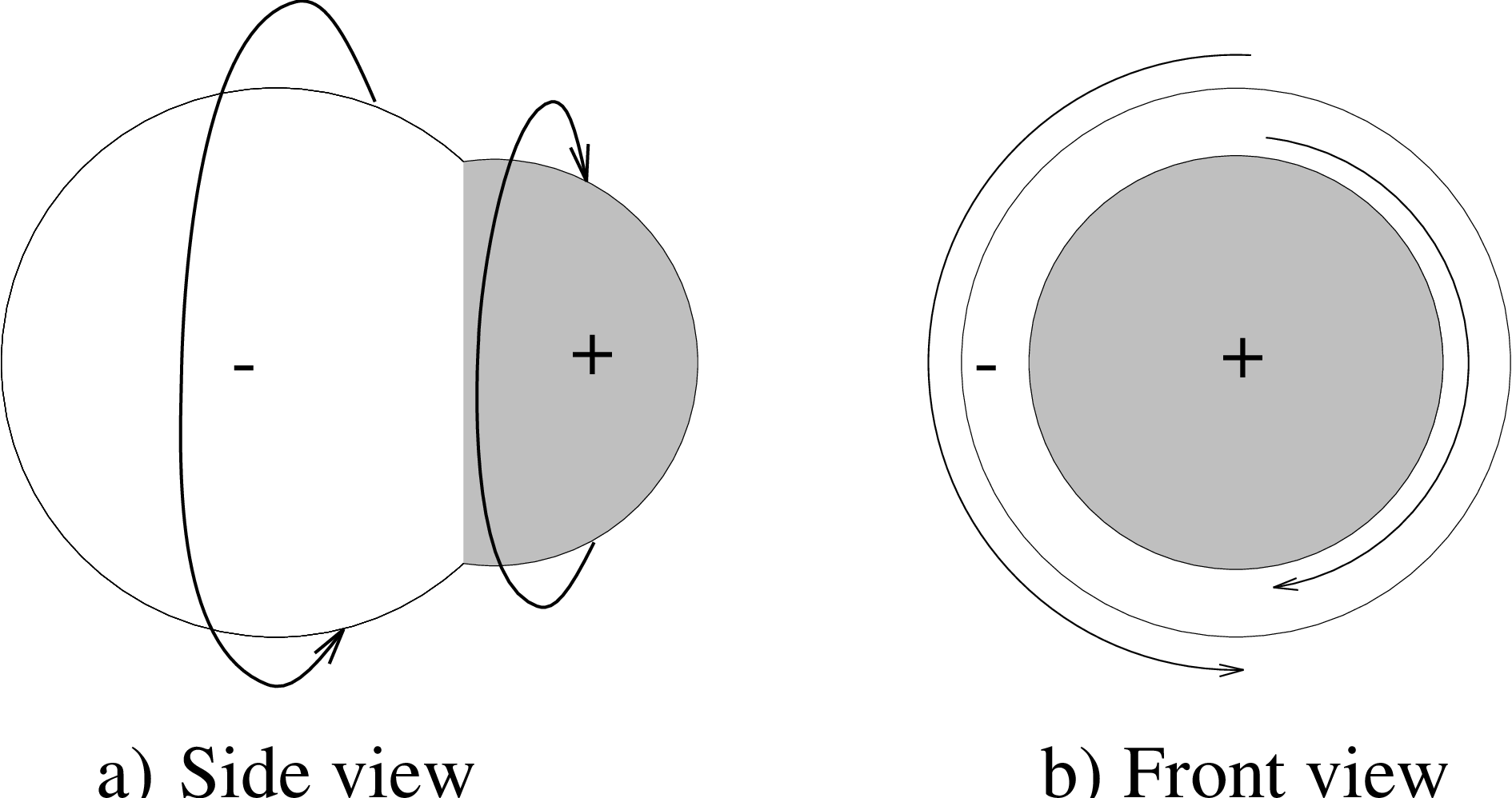}
\caption{The superpair as a single particle system with two members of opposite sign and spin.}
\label{minipair2}
\end{figure}

The conventional picture that a kind of pairing occurs between {\em{two normal}} electrons mediated by `something' to form the superelectron pair is a right conclusion but may be with  a wrong picture that has led the theory to  difficulties in explaining how to overcome the Coulomb repulsion force in HTS.  The origin of the pair may in fact  be the electron itself deformed (`splitted') under the applied cooling conditions to a ``self-paired'' superelectron.  This superpair ought not to be considered either as pure opposite electric charges or as  pure tiny magnet but a combination of both.  It is a single dynamical system with two members of opposite sign, and spin originated from within the normal single electron  as illustrated in Fig. \ref{minipair2}.

The superpair free motion would ``stop'' and the D. C.   current would flow without free movement through out the crystal. This may occurs in some way with "pulsating" or "transferring" the energy by the collective coherent coupling with no resistance, i.e.  the superpair is no longer the familiar charge carrier, but a charge transferor.  If any energy above a certain value forces it to move freely (individually), it will be restored again to a normal electron charge carrier which we are familiar.  This process will be described below with more detail.

This specific transmutation, occurring with the  conduction electrons, might be closely related to the nuclei of the atoms.  Simultaneously with this transmutation of the conduction electrons, another specific transmutation occurs with a corresponding number of neutrons in the nuclei.  Every such  neutron begins to exhibit apparent properties that allow it to interact with a corresponding proton forming another type of pair system. This internal pair would interact with the outer paired electron in a complicated system to keep the balance and symmetry in the atomic system.  What is meant by the idea of electron core? How all these transmutations occur in the sense of the proposed picture and how are they related in the atomic system?
We understand the electron core is that region of space where the electron cloud is densified and bent around it as visualized in Fig. \ref{cornorm}.

The system \{Electron Core-Cloud\}-\{Nucleus\} interaction can be considered as  a system of \{Electron Core-Cloud\}-\{Proton-Neutron\}, so it can be divided into two directly related subsystems:
\{Electron Core-Neutron\} and \{Electron Cloud-Proton\}.
\begin{figure}[ht]
\centering
\includegraphics[width=0.75\textwidth]{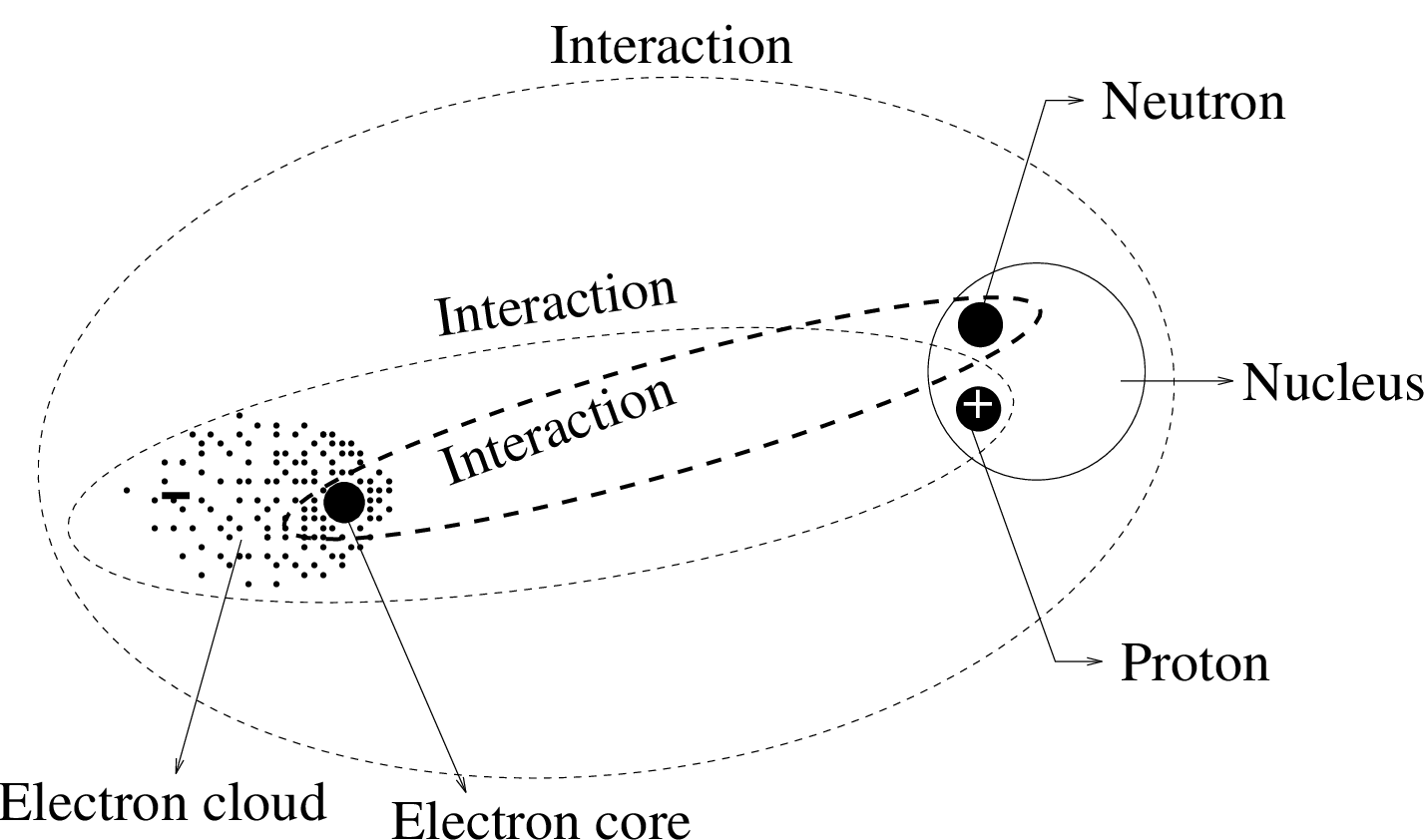}
\caption{Diagram of the system interaction in the normal state.  The heart of this system is the Electron Core-Neutron interaction.}
\label{diagnor}
\end{figure}
\begin{figure}[ht]
\centering
\includegraphics[width=0.75\textwidth]{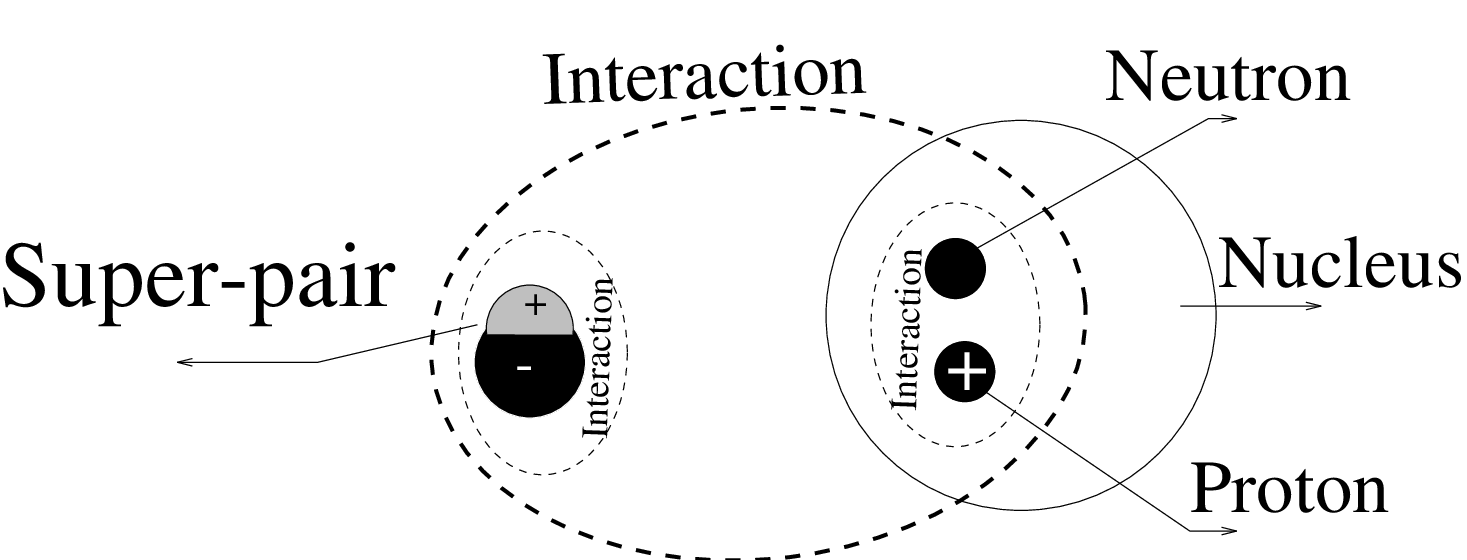}
\caption{Diagram of the system interaction in the superconducting state.  The heart of this system is the \{Super-pair\}-\{Proton-Neutron\} interaction.}
\label{diagsup}
\end{figure}
These subsystems interact within one  unified atomic system in different states as follows:\\
\begin{enumerate}
\item The normal  state.  The heart of the system \{Electron Core-Cloud\}-\{Proton-Neutron\} is the
\{Electron Core-Neutron\} interaction as shown in the diagram of  Fig. \ref{diagnor}.\\
\item The superconducting state.  The \{Superpair\}-\{Proton-Neutron\} interaction is the heart of the system as illustrated in Fig. \ref{diagsup}
\end{enumerate}
\begin{figure}[ht]
\centering
\includegraphics[width=0.6\textwidth]{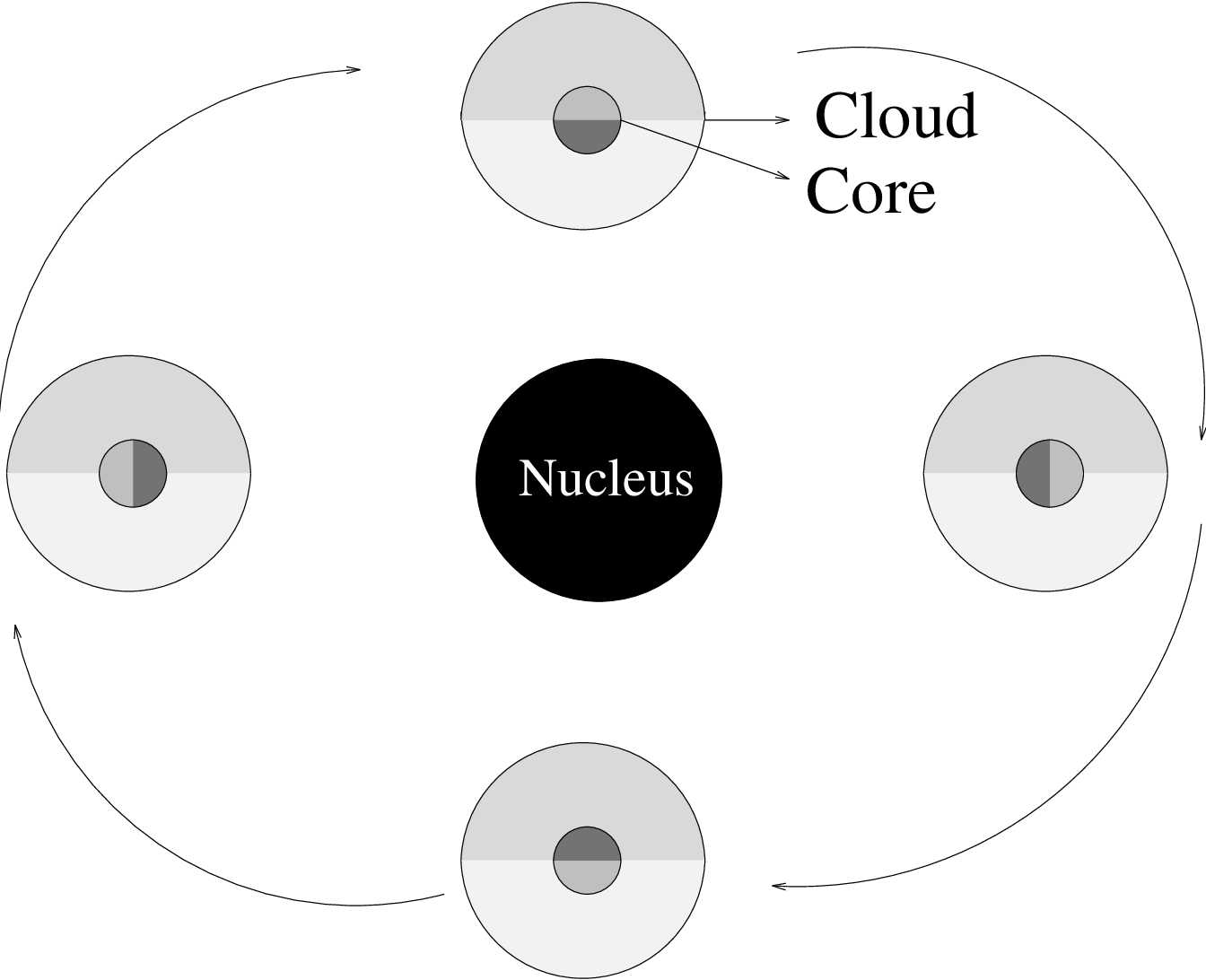}
\caption{The rotation of a normal electron ($T>>T_c$) with its core around the nucleus.  One half of each is highlighted to enable observing their relative
 rotations. The rotation centre of the system Electron Core-Cloud lies in the Proton-Neutron system.}
\label{rotatnor}
\end{figure}
\begin{figure}[ht]
\centering
\includegraphics[width=0.6\textwidth]{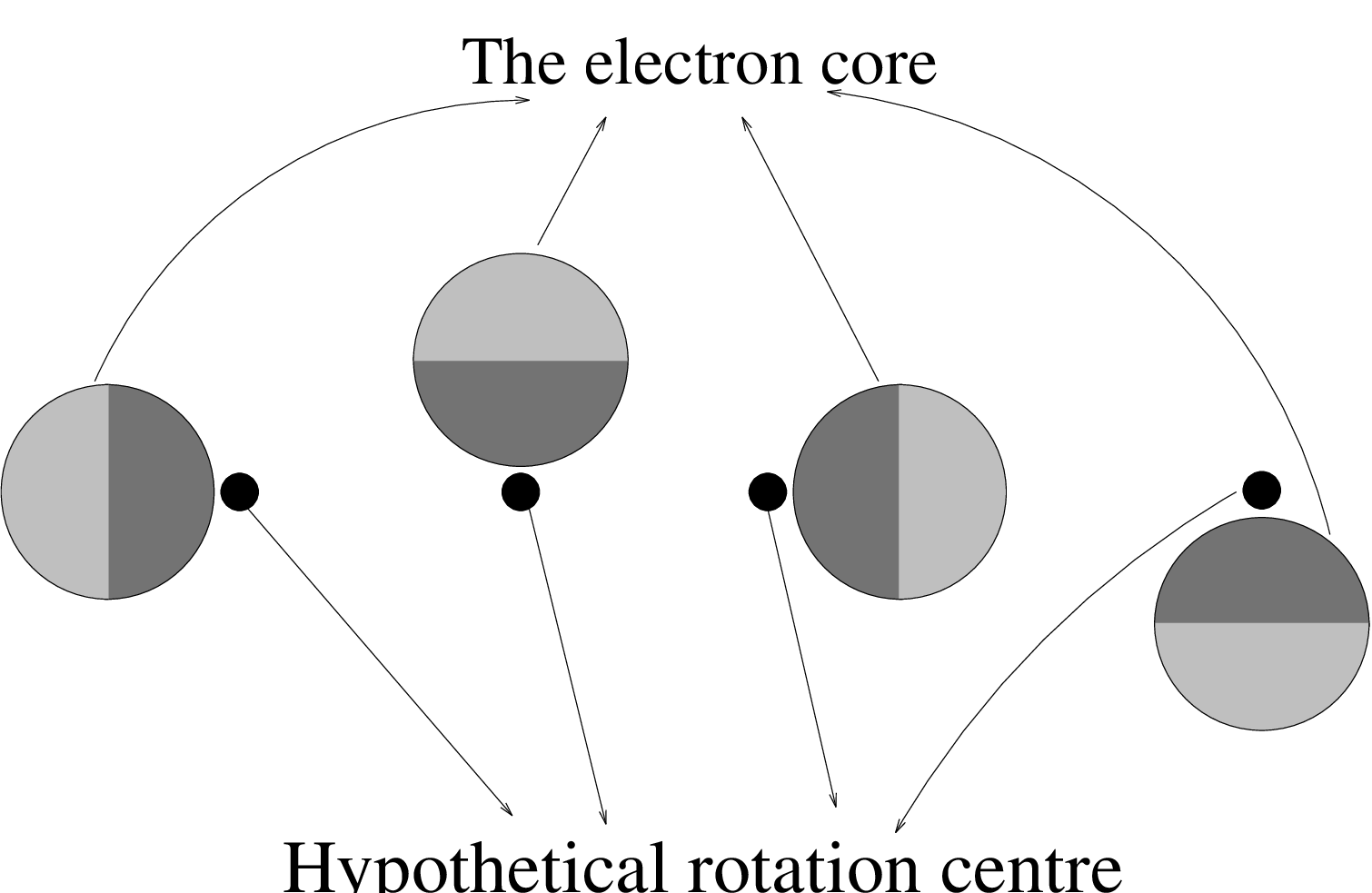}
\caption{The rotation change of the core at the beginning of the splitting process (growing of the core and collapsing the cloud ($T>T_c$).  The rotation centre (hypothetically) is moved closer to the core.}
\label{rotatnor1}
\end{figure}
\begin{figure}[ht]
\centering
\includegraphics[width=0.6\textwidth]{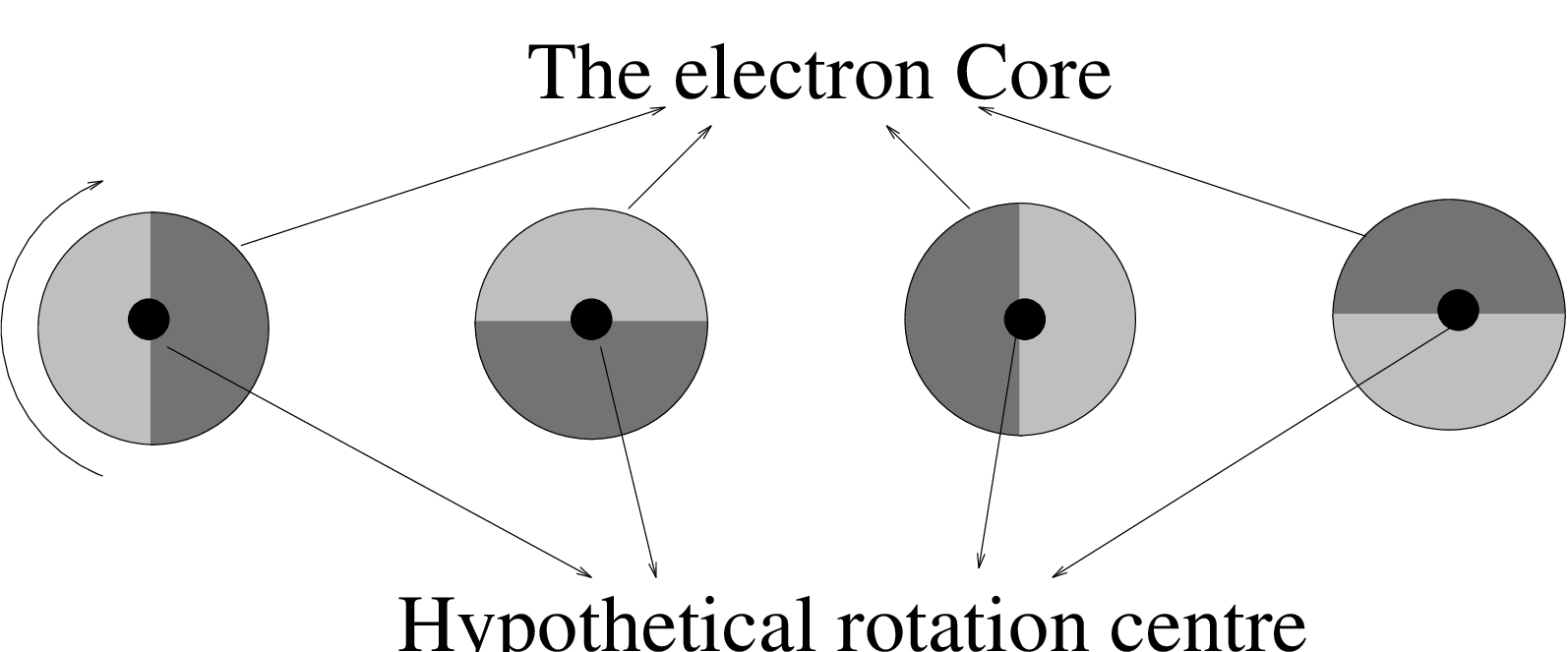}
\caption{The rotation change of the core at and below the transition ($T\leq T_c$). The rotation centre is moved inside the superpair.}
\label{rotatsup}
\end{figure}
The process of the electron splitting is initiated when the system   described in the normal state (see Fig. \ref{diagnor}) is disturbed, i.e.  when the electron cloud starts contracting and the core expanding. Although the core would still be `screened' above  $T_c$  (see Fig. \ref{minipair}), the properties of a given material would be affected from the beginning of this splitting process (sufficiently above $T_c$).  However, why would the electron core have a
dual character and how  does it change to exhibit the apparent properties of a positive charge after breaking through the cloud at $T_c$ as shown in Fig. \ref{minipair2}?

Three phases of the electron core may be distinguished:

\begin{enumerate}
\item In the normal state (before the beginning of the splitting process).  The rotation of a normal electron around the nucleus is illustrated in Fig. \ref{rotatnor}. For visual convenience the cloud and the core of the same electron at different locations of the orbit are divided into two faces to observe their relative rotations. The core is drawn relatively large for convenience also.  The core is always with the same face directed towards the nucleus, while the cloud is with different faces as clearly shown in this figure.  The rotation centre of the system \{Electron Core-Cloud\} lies in the \{Proton-Neutron\} system.

\item In the normal state (at the beginning of the splitting process but before the transition).  In this case the rotation (hypothetical) centre begins to move stepwise towards the core as shown in Fig. \ref{rotatnor1} where only the core faces are shown relative to the centre.

\item At the transition temperature (the electron is splitted to a superpair(see Fig. \ref{minipair2})), the rotation centre is moved inside the superpair 
as shown in Fig. \ref{rotatsup}. Again the core faces only are shown relative to the centre.
\end{enumerate}

By this way the electron core at the transition temperature apparently changes its properties to that visualized in Fig. \ref{minipair2}. This namely  implies that the free motion of the superpair would ``freeze'' and vice versa, if we manage to ``freeze'' the free motion of the normal electron, it would split to a superpair.

So the interaction systems may be summarized in two distinct phase transitions as:

1.  In the normal  state the system is divided into two subsystems:
\begin{itemize}
\item Electron Core-Neutron
\item Electron Cloud-Proton
\end{itemize}

2.  In the superconducting state, the system is (external pair-internal pair) and contains two subsystems:
\begin{itemize}
\item Electron Core-Cloud, i.e.  the external pair (superpair)
\item Proton-Neutron, i.e. the internal pair
\end{itemize}

Any disturbance to the electron superpair will be reflected to the internal pair (former Neutron-Proton).  This would suggest that the outermost conduction electrons of a given material should be sufficiently bound to the  nuclei of the atoms to favour the pairing at higher temperatures.  This could be also one of the reasons that a good conductor (the conduction electrons are very weakly bound to their nuclei) is difficult to make a superconductor if not at al.

\section{New Insight into Some Important Evidences}

\subsection{The D.C. Current in a Superconductor}
One may question how a D. C.  current would flow without  charge carriers (normal electrons) inside a superconductor connected to  a D. C.   voltage via normal conductors?  Schematically, the picture in general analogy, may be described as follows:

By deforming (splitting) the conduction electrons to superpairs of opposite sign members, all superpairs would be coupled coherently via their specific intrinsic fields. By means of this coherent coupling, the current can be transmitted from one to the next pair.  But in what way?

\begin{figure}[ht]
\centering
\includegraphics[width=0.75\textwidth]{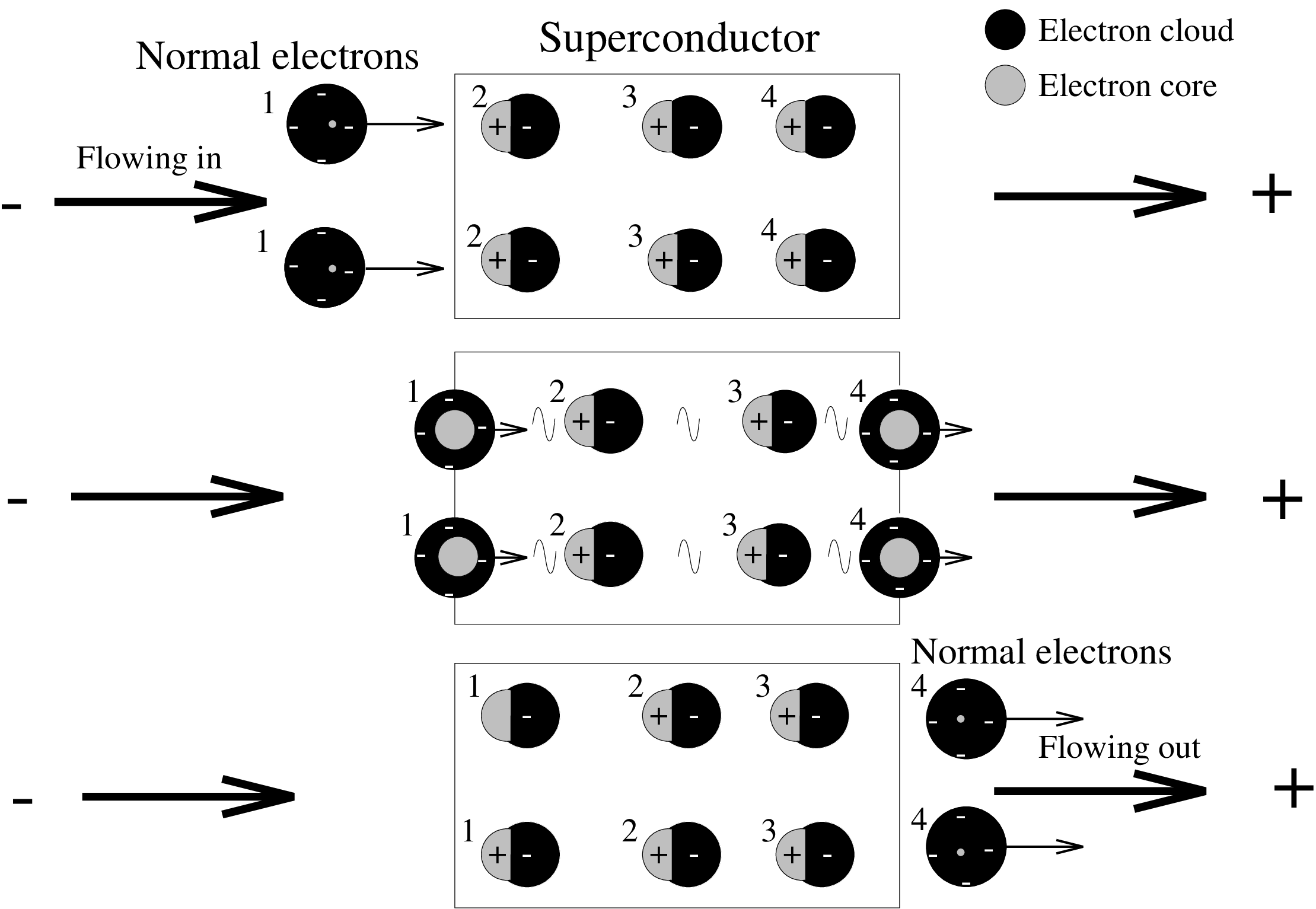}
\caption{The flow of D. C.  current through a superconductor connected to a D. C.  voltage.}
\label{dccurrent}
\end{figure}

The normal electrons flowing from the lower towards the higher potential of the source, jump at the contact junction inside the superconductor and split to superpairs.  Their kinetic energies and abrupt transmutation into superpairs cause a kind of wave to the local coupled superpairs and they coherently pulsate this wave along  the superconductor.  At the same time, a tiny coherent collective displacement occurs (without any `friction/colisions' with the lattice or with each other because they no longer move individually free as in the conventional conductors) which allows the new paired electrons to settle and force equal portion of pairs at the other end of the superconductor to be pushed out of it as shown in Fig. \ref{dccurrent}.

\begin{figure}[ht]
\centering
\includegraphics[width=0.6\textwidth]{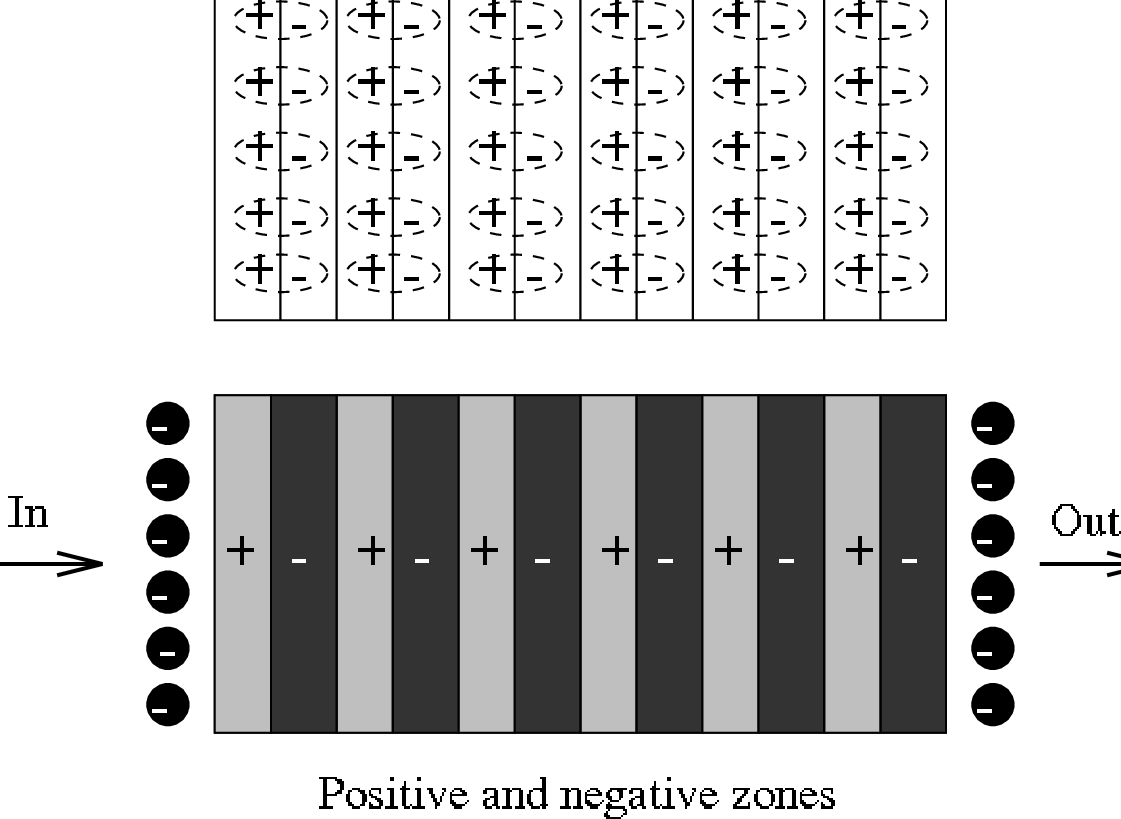}
\caption{Negative and positive zones produced in a superconductor by paired electrons}
\label{dczones}
\end{figure}

These expelled pairs transmute immediately to normal electrons with the same energy of those normal electrons just before their jump in the superconductor.  So we observe the same current flowing in and flowing out of the superconductor with zero resistance (no individual scattering/drift of paired electrons) and much smoother than in a normal metal. The paired and coupled electrons produce kind of negative and positive zones as illustrated in Fig. \ref{dczones}. Every zone has the same properties of the corresponding member of the pair. The displacement occurs with the whole zones.


\subsection{Josephson Effects}

Josephson effects and their applicability have turned to be  nontrivial problems because the physical picture is still unobvious in many observations considering the existing pairing mechanisms.  These effects also are capable to support strongly the proposed picture, i.e.   the existence of self electron superpair of opposite sign members in the superconducting state.

\subsubsection{The D.C Josephson Effect}
In the case of no applied external voltage to a junction made of  superconductor-insulator-superconductor (S/I/S) , the following process may be described:
\begin{figure}[ht]
\centering
\includegraphics[width=0.6\textwidth]{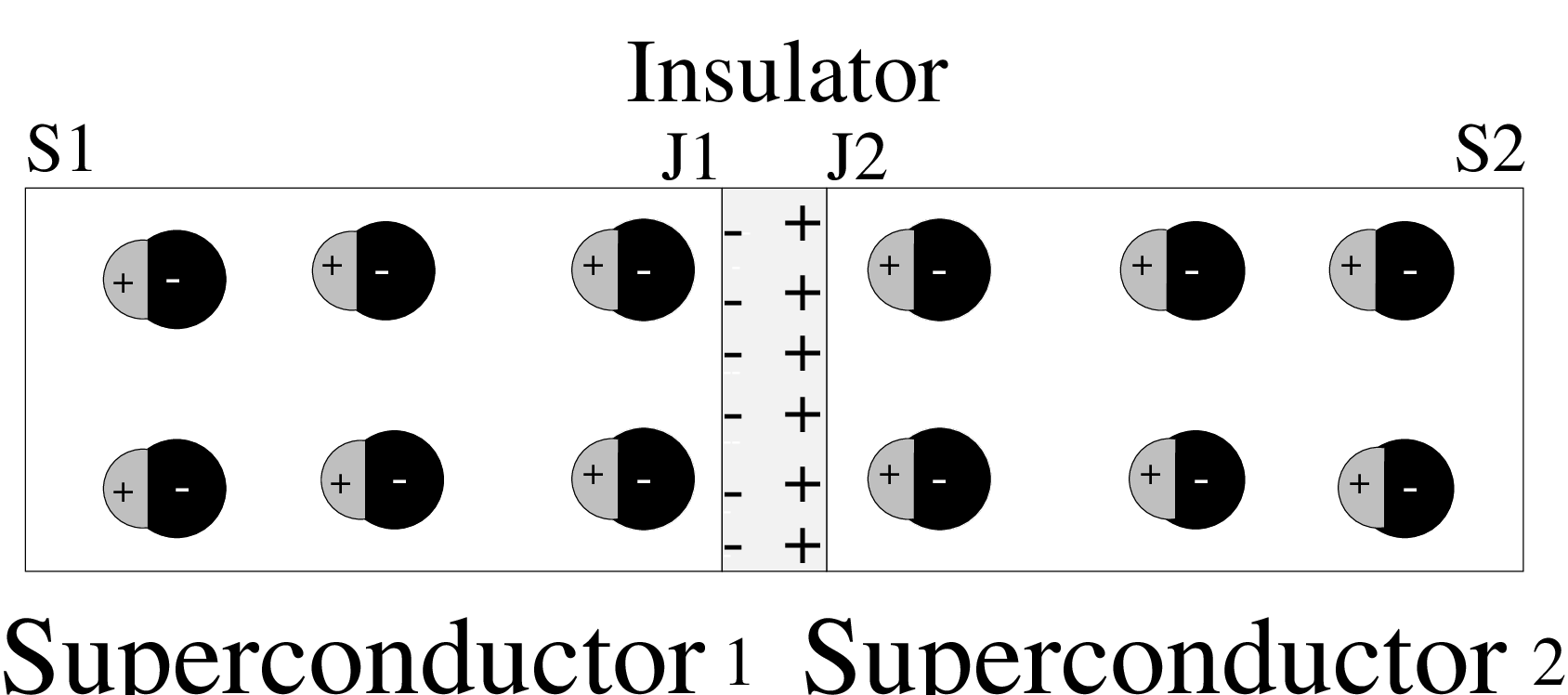}
\caption{The D. C.  effect with no applied external voltage.}
\label{dcjunc}
\end{figure}
This picture makes us to suggest that the superpairs do not tunnel through the barrier.  Only restored pairs (normal electrons) would tunnel throw the junction.  The negative and positive zones oriented at the edges of the junction would interact with each other.  They tend to join both sides of the superconductor to achieve the coherent coupling as in the rest of  the ``previously aligned'' superconductor.  This causes the appearance of some potential at the edges of the junction as shown in Fig. \ref{dcjunc}.

\subsubsection{The A.C Josephson Effect}
The A. C.  Josephson effect seems to be more complicated and maybe explained as follows:

By applying a small D. C.   voltage at the outer edges (say $S_1$ connected to  the positive  and $S_2$ with the negative potential) of two superconductors separated  by sufficiently thin insulator at the junction edges denoted by $J_1$ and $J_2$,  the positive potential of the voltage tends to give a total positive charge to the first  superconductor, i.e.  ``+'' to its discontinuities $S_1$ and $J_1$.  The negative potential tends the  other way, i.e.  to give ``-'' to $S_2$ and $J_2$ of the second superconductor. If the junction separates two normal conductors this is the case, but the current would flow only after breaking down the insulator with discharge caused by  a relatively high break down voltage. If there is no junction separating the superconductor we refer  to the picture described in Fig. \ref{dccurrent}

\begin{figure}[ht]
\centering
\includegraphics[width=0.75\textwidth]{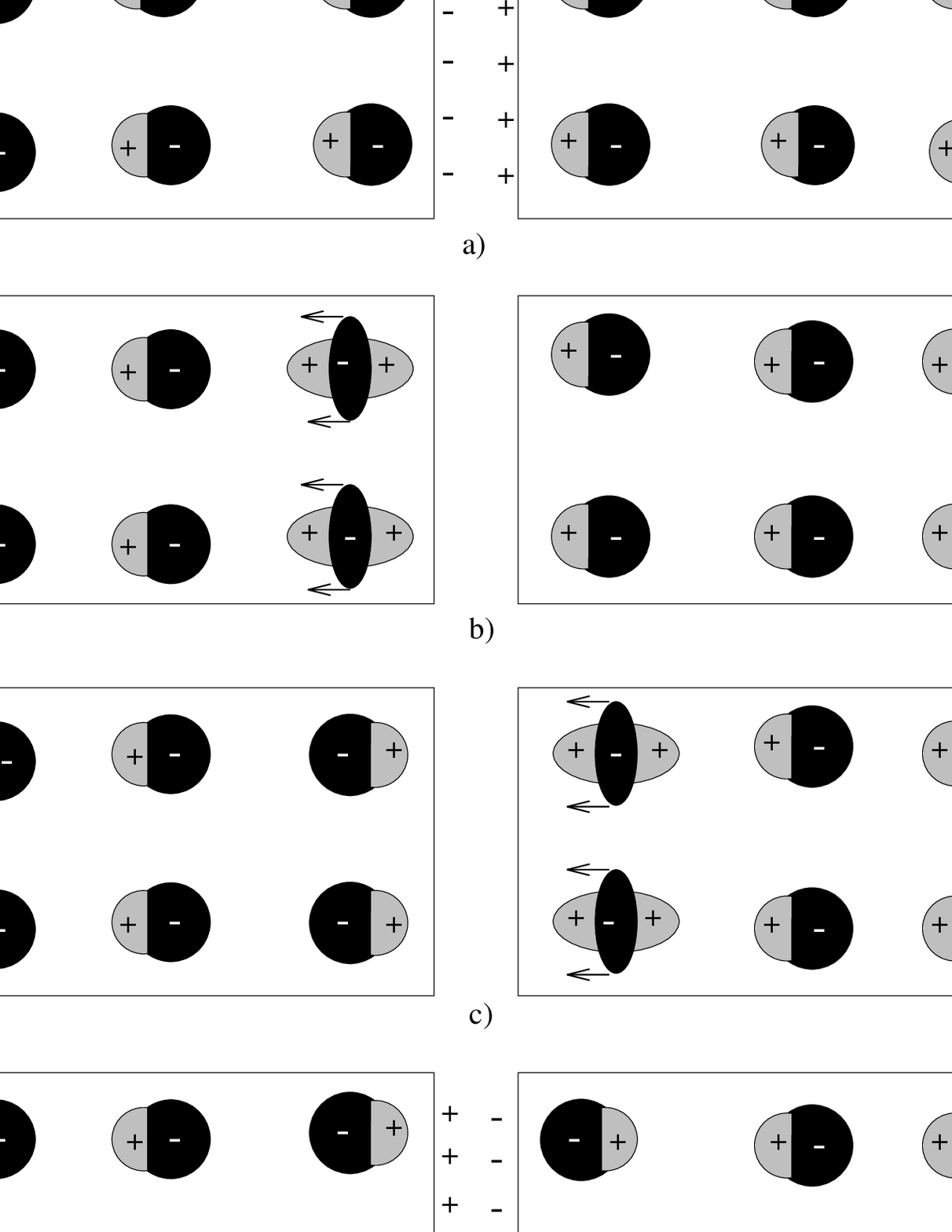}
\caption{The D. C.  effect with no applied external voltage.}
\label{acjunc}
\end{figure}

In  the case of two separated superconductors, however, the introduced structure of the superpairs prevents this situation.  They tend  $S_1$ to be with "+" and $J_1$ with ``-'' as a result of  the pairing symmetry (see Fig. \ref{acjunc} a).  By analogous manner $S_2$ would be with ``-'' and $J_2$ with ``+''.  Only at the junction edges, an instability situation occurs. It forces the superpairs to experience further deformation. The electron clouds (the members  of the pairs with sign ``-'') at $J_1$  begin to ``slide'' backward  uncovering the other members (with sign ``+'') from the junction side (Fig.\ref{acjunc} b).  The opposite occurs at $J_2$  as shown in Fig. \ref{acjunc} c, d.  The first half period is completed.
 Now $J_1$ is already positive and $J_2$ is negative (Fig.\ref{acjunc} d).  Instability situation arises between the pairs at the edges of the junction and the neighbouring internal pairs because of repulsive forces (see Fig. \ref{acjunc} d).  The negative clouds of the pairs at  $J_1$  ``slide'' again towards the edge $J_1$ uncovering the positive core  from the internal side of the superconductor. Finally the same arrangement shown in Fig. \ref{acjunc} a is achieved (after a complete period).   The same process occurs at $J_2$ with opposite sign and so on.   The vibration of the pairs at the junction edges will be maintained producing A. C.  voltage as long as the D. C.  voltage is applied. Possibly this would explain also the phase shift of $\pi$ observed in many experiments at such  junctions.

\section{Conclusions and Suggestions}

In summary, any free movement of charge carriers is always accompanied with ``friction/colisions'' which increase the heat throughout the conductor and consequently the resistance.  Once the electron free motion is eliminated, it deforms or `split' to a specific superpair structure with a new function of transferring the energy, but no longer a charge carrier.  Any applied external energy overcomes and breaks this condition and forces it to move individually, it will be restored again to a normal electron charge carrier.  No pairing of electrons could be  achieved if not accompanied by a corresponding change in the nuclei and vice versa.  The cooling to a certain temperature may be only one of other means to achieve and maintain (conserve) the electron pairing.

Concentrating the research and experimental efforts for investigating the probably more complicated structures of the electrons, we may could explore the dual character of their cores.  Probably we could find also that the change with the electrons should be accompanied by a corresponding change with some particles in the nuclei (probably the neutrons) of the atoms.  This on the other hand could certainly be helpful to explain why superconductivity occurs with different elements, alloys and compounds at different transition temperatures. Why isotopic substitution, ``carrier'' concentration, tension,  pressure and many other factors  shift the transition temperature up or down. Why a good conductor is difficult to make a superconductor if not at al.

World wide extensive experimental work may be directed on search of  methods how to conserve the superpairs artificially to maintain their structure at much higher temperatures above $T_c$.  Another essential experimental direction is to find methods how to favour the electron pairing (without conventional cooling) by changing the function of some neutrons to interact with some protons forming the above mentioned internal pairs.  This would help us also to find the right way to prepare an alloy from more stable and reliable probably semi-metal materials to exhibit superconductivity, hopefully, at room temperature and ambient presure. A preliminary suggestion is a combination of Magnesium-Niobium-Strontium.   

Although the new proposed physical picture of the origin and structure of the pairing is strange at first, it would, however, eliminate the problem of the strong Coulomb repulsion forces at high temperature as the proposed pair structure turns the forces naturally to attractive forces tightening the coupling between all the pairs to behave collectively.

It should be emphasized, however, that this assumption of the electron structure  would not turn down any established fundamental law in physics. The  old atomic models of protons and neutrons dominated for long years without knowing about their composites, the quarks. The puzzling superconducting phenomena encourage
such asumption as a possibility which needs to be tested  and justified sufficiently with unambigious evidences. It is far from complete and thus it is called from the begining as just an initial idea. Unlike mathematical models in physics, here the difference is that a physical model is developed upon a physical assumption to explain well known effects.

\end{document}